\documentclass[aps,prl,reprint]{revtex4-1}
\usepackage[utf8]{inputenc}
\usepackage[T1]{fontenc}
\usepackage{newtxtext, newtxmath} 
\usepackage{graphicx, braket, siunitx, mathtools, hyperref}

\let\Im\undefined\DeclareMathOperator\Im{Im}
\DeclarePairedDelimiter\av\{\} 

\begin{document}

\title{Strong-coupling quantum logic of trapped ions}
\author{Mahdi Sameti}
\affiliation{Blackett Laboratory, Imperial College London, London SW7 2AZ, United Kingdom}
\author{Jake Lishman}
\affiliation{Blackett Laboratory, Imperial College London, London SW7 2AZ, United Kingdom}
\author{Florian Mintert}
\affiliation{Blackett Laboratory, Imperial College London, London SW7 2AZ, United Kingdom}

\date{\today}

\begin{abstract}
Essentially all known quantum gates rely on a weak-coupling approximation resulting in linear dynamics.
With the explicit example of trapped ions, we show how high-fidelity quantum gates can be achieved outside such an approximation, and we derive readily implementable driving fields to realize gates with extremely high fidelities for ions well outside the Lamb--Dicke regime with motional temperatures achievable by only Doppler cooling.
\end{abstract}

\maketitle

Entangling quantum gates are the central element in quantum information processing.
After decades of experimental effort, such gates have successfully been realized in several physical systems including trapped ions~\cite{Schaefer18,Harty16,Gaebler16}, superconducting circuits~\cite{Reagor18}, quantum dots~\cite{Veldhorst15} and NV centers~\cite{Rong2015}.
After a period of proof-of-principle experiments, the field now requires fast quantum gates with extremely high fidelities for the next step towards hardware that can outperform classical devices.

Among the most advanced platforms are trapped ions~\cite{Bermudez17,Bruzewicz19}.
Since ions are spatially separated due to their Coulomb repulsion, there is no appreciable direct interaction between the electronic degrees of freedom that define the qubits, and effective interactions mediated via collective motional modes need to be engineered in order to realize entangling gates.
This mechanism involves a change in the motional state~\cite{Leibfried03} that is absolutely essential for the implementation of the gate.
It is, however, equally essential that the electronic and motional modes become uncorrelated at the gate time, as to do otherwise would result in an incoherent gate operation.

There are a variety of schemes~\cite{Sorensen00,Garcia05,Ospelkaus08,Bermudez12} to drive ions with electromagnetic fields that achieve this in the \emph{Lamb--Dicke regime} of weak ion--motion interactions with the motional modes all at low temperatures.
For most of the currently employed entangling gates in the Lamb--Dicke regime, comparably simple driving schemes result in gate operations that are largely independent of the initial motional state.

Being restricted to the Lamb--Dicke regime nonetheless imposes several challenges.
Due to the weak interactions, realizations of fast gates require strong laser driving causing adverse effects like AC Stark shifts and off-resonant excitations of undesired transitions, which lower the gate fidelity~\cite{Steane00}.
The necessity to cool ions close to the ground state implies that only a limited number of gates can be performed between cooling cycles, which decreases the number of gates that can be executed within the coherence time.
Even with perfectly cooled motion and weak interactions, the Lamb--Dicke approximation still fails to be sufficient in the quest for entangling gates of ever-higher fidelity~\cite{Schaefer18,Harty16,Gaebler16}.
The present goal is thus to devise an entangling quantum gate for trapped ions that can be realized well outside the validity of the Lamb--Dicke approximation.

\begin{figure}
    \includegraphics{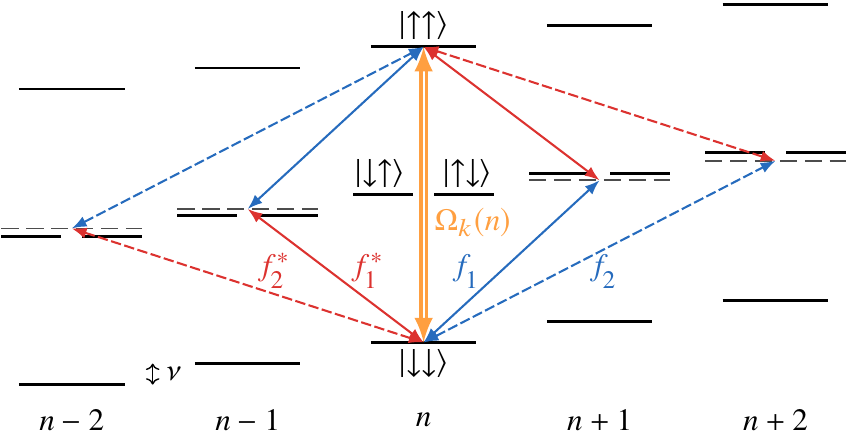}
    \caption{\label{fig:diagram}%
Energy level diagram for two ions and one motional mode.
Electronic (qubit) levels of ions are denoted by $\ket{\downarrow}$ and $\ket{\uparrow}$, and the motional state is characterized by the phonon number $n$.
Driving a $k$-th order blue and red sideband transition simultaneously close to resonance results in an effective qubit--qubit interaction with coupling constant $\Omega_k$.
Outside the Lamb--Dicke regime this coupling depends on the phonon number $n$.}
\end{figure}

The construction will be exemplified on the M\o lmer--S\o rensen gate~\cite{Sorensen99,Sorensen00}, but applies equally to the full range of similar gates~\cite{Roos08,Cohen15,Haljan05} that are currently used.
The basic principle of the entangling gate can be appreciated using the level diagram in Fig.~\ref{fig:diagram}.
In addition to carrier transitions with no change in the motional mode, there are also sideband transitions; in a $k$-th order red (blue) sideband transition an excitation/de-excitation of an ion is accompanied by the annihilation/creation (creation/annihilation) of $k$ phonons.
First-order blue and red sideband transitions are depicted by solid blue and red arrows, and second-order sideband transitions are depicted by dashed arrows.
The gate relies on the simultaneous driving of both the red and blue first-order sideband transitions close to, but not exactly on, resonance.
Apart from spurious excitations and de-excitations of phonons that can be made to vanish at the end of the gate operation, this results in an effective interaction between the electronic degrees of freedom, as depicted by an orange (thick) arrow in Fig.~\ref{fig:diagram}.

This is a truly coherent qubit--qubit interaction only under the Lamb--Dicke approximation; without it, the effective interaction strength becomes dependent on the initial state of the phonon mode and a coherent gate operation can only be ensured if this is originally a Fock state.
Realistically, the motional state will be a statistical mixture of several Fock states, such as a thermal state, and the gate operation will become incoherent.

As will be shown here, coherent gates outside the Lamb--Dicke regime can be realized in terms of suitable driving of higher-order sideband transitions, and the driving profiles can be obtained based on a systematic expansion in terms of the Lamb--Dicke parameter $\eta$ that characterizes the coupling between electronic and motional degrees of freedom of the ions.

The Hamiltonian in the interaction picture for trapped ions and one motional mode driven with an external laser or microwave field with time dependence $f(t)$ reads~\cite{Roos08}
\begin{equation}\label{eq:interaction}
H = f(t) S_+ \exp\Bigl(i \eta\bigl(a e^{-i \nu t} + a^\dagger e^{i \nu t}\bigr)\Bigr) + \text{H.c.}\,,
\end{equation}
where $\nu$ is the frequency of the motional mode; the Lamb--Dicke parameter $\eta$ is given by the ratio of the photon momentum of the driving field to the phonon momentum of the motional mode; $S_\pm = \sum_j \sigma_\pm^{(j)}$ are the collective qubit raising and lowering operators; and $a$ ($a^\dagger$) are the phonon annihilation (creation) operators.

All the previously mentioned complications arising outside the Lamb--Dicke regime result from the non-linear dependence of $H$ on the phonon creation and annihilation operators.
In order to appreciate this, it is instructive to express the exponential function in Eq.~\eqref{eq:interaction} as
\begin{equation}\label{eq:preDk}
\exp\Bigl(i \eta \bigl(a e^{-i\nu t} + a^\dagger e^{i\nu t} \bigr)\Bigr)
= e^{-\eta^2/2} \sum_{k=-\infty}^\infty{ D}_{k}(\eta) e^{ik\nu t}\,,
\end{equation}
with
\begin{equation}\label{eq:Dk}
\mathcal D_{k}(\eta) = \sum_{n=0}^{\infty} (i\eta)^{2n+k}\frac{{a^{\dagger}}^{n+k}}{(n+k)!} \frac{a^n}{n!} \quad \text{for $k \ge 0\,,$}
\end{equation}
and $\mathcal{D}_{-k}(\eta) = \mathcal D_{k}^\dagger(-\eta)$.
The term $\mathcal{D}_{0}$ corresponds to carrier transitions since it preserves the phonon number, and the terms $\mathcal{D}_{k}$ with $k>0$ and $k<0$ correspond to blue and red sideband transitions of order $k$ respectively.
In the frequently employed Lamb--Dicke approximation, the sum in Eq.~\eqref{eq:Dk} is restricted to terms that are at most first order in $\eta$, resulting in a linear Hamiltonian corresponding to linear Heisenberg equations of motion.
In general, however, the qubit--phonon coupling is non-linear as reflected by Eq.~\eqref{eq:Dk}.

Most entangling gates are realized in terms of a two-photon process comprised of a transition of both the first-order red and blue sidebands, with effective coupling constant
\begin{equation}\label{eq:Weff}
\Omega_1\propto \bigl[\mathcal D_{1}(\eta),\,\mathcal D_{-1}(\eta)\bigr]= \eta^{2} - 2\eta^4 a^\dagger a + \mathcal{O}(\eta^6)
\end{equation}
for the entangling qubit--qubit interaction.
In lowest order ($\propto\eta^2$) this is indeed independent of the initial phonon occupation, whereas the dependence on $a^\dagger a$ at fourth and higher orders modifies the effective process from a coherent interaction between two qubits to a three-body interaction between two qubits and the motional mode.

Driving additional blue and red sidebands of the same order simultaneously engenders complementary resonant two-photon processes which contribute additional terms $\Omega_k\propto [\mathcal D_{k},\,\mathcal D_{-k}]$ for $k>1$ to the coupling constant.
The goal of the present approach is to combine simultaneous driving of sufficiently many higher-order sidebands with appropriate fields such that the phonon-number-dependent processes cancel.

To achieve this, we consider driving protocols with the generic temporal pattern
\begin{equation}\label{eq:driving-field}
f(t) = -i\frac{e^{\eta^2/2}}\eta\sum_{k>0}\Bigl(f_k(t)e^{-i k \nu t}+{(-1)}^kf_k^*(t)e^{i k \nu t}\Bigr)\,.
\end{equation}
The first (second) term in this ansatz corresponds to the driving of the $k$-th order blue (red) sidebands, and the factors $f_k(t)$ vary slowly in time to ensure that this driving is slightly off-resonant.
The phase factor ${(-1)}^k$ and the prefactor $e^{\eta^2/2}\eta^{-1}$ can be understood as convention that can be chosen at will, as long as the factors $f_k(t)$ are not determined yet.
This particular choice of phase factors will result in a more systematic expansion later on, and the factor $e^{\eta^2/2}$ is chosen to cancel the first factor in Eq.~\eqref{eq:preDk}.
While these choices are mostly for convenience, the prefactor $\eta^{-1}$ is essential for the expansion in powers of $\eta$ and it reflects the fact that a decreasing Lamb--Dicke parameter requires increasing amplitudes of the driving fields in order to maintain a constant entangling interaction.

Neglecting far-off-resonant processes in the interaction Hamiltonian of Eq.~\eqref{eq:interaction} with the explicit driving profile in Eq.~\eqref{eq:driving-field} results in the compact Hamiltonian
\begin{equation} \label{eq:generalized-hamiltonian}
H_{s} =  \frac1\eta S_y\sum_{k}\Bigl(f_k(t)  \mathcal D_k(\eta) + f_k^*(t)  \mathcal D^\dagger_{k}(\eta)\Bigr)\,.
\end{equation}
Inside the Lamb--Dicke regime this reduces to the linear Hamiltonian $H_0 = iS_y(f_1a^\dagger - f_1^*a)$, for which the time evolution is given by
\begin{equation}\label{eq:propagator-0}
\mathcal U_{0}(t) = \exp\Bigl(S_y \bigl(\av{f_1}a^{\dagger} - \av{f_1^*}a\bigr)
                               + i\Phi_0 S_y^2\Bigr)\,,
\end{equation}
with the Rabi angle $\Phi_0(t)=\Im\av[\big]{f_1 \av{f_1^*}}$ specified in terms of the shorthand notation $\av{f}=\int_0^t\mathrm{d}t_1\,f(t_1)$ and its nesting $\av[\big]{f\av{g}}=\int_0^t\mathrm{d}t_1\,f(t_1)\int_0^{t_1}\mathrm{d}t_2\,g(t_2)$.

In order to solve the system dynamics outside the Lamb--Dicke regime including terms up to a high order in $\eta$, the exact propagator $\mathcal U$ is approximated by a product $\mathcal V_d$ of time-dependent unitaries $\mathcal U_j$ as
\begin{equation}\label{eq:product}
\mathcal U \approx \mathcal{V}_d = {\cal U}_0{\cal U}_1\dotsm{\cal U}_{d-1}{\cal U}_d\,,
\end{equation}
so that the transformed Hamiltonian
\begin{equation}
H_{j+1}=\mathcal{V}_{j}^\dagger H_s\mathcal{V}_{j}-i\mathcal{V}_{j}^\dagger\dot{\mathcal{V}}_{j}
\label{eq:Hj}
\end{equation}
contains only terms of order at least $\eta^{j+1}$.
With the solution of the linearized problem $\mathcal{U}_{0}$, this is naturally ensured for $H_1$.
Denoting the terms in leading order of $\eta$ of $H_j$ by $\bar{H}_j$, \textit{i.e.}\ $\bar H_j\propto \eta^j$, one can define the propagator $\mathcal{U}_{1}=\exp\bigl(-i\av{\bar{H}_1}\bigr)$.
Since $\mathcal{U}_{1}$ solves the Schr\"odinger equation with Hamiltonian $H_1$ in leading order of $\eta$, $H_2$ is of order at least $\eta^2$.
This process can be iterated by defining the unitary $\mathcal{U}_j$ entering Eq.~\eqref{eq:product} as $\mathcal{U}_j(t) = \exp\bigl(-i \av{\bar{H}_j}\bigr)$.
This definition ensures that the subsequent Hamiltonian $H_{j+1}$ defined in Eq.~\eqref{eq:Hj} contains only terms of order at least $\eta^{j+1}$.

After $d$ steps one arrives at a Hamiltonian $H_{d} \sim \eta^d$ which still permits an exact solution to the Schr\"odinger equation, but at this step it is approximated by $\mathcal{U}_{d}=\exp(-i\av{H_d})$, where in contrast to the previous steps, this construction is not limited to the dominant part of $H_d$.
This approximate propagator differs from the exact propagator by terms of order at least $\eta^{2d}$ since the lowest-order terms that are being neglected are bilinear in $H_d$.
This systematic construction thus yields an approximation of the actual propagator that is accurate to high orders in $\eta$
and is an ideal starting point for the design of the desired driving fields.

The desired entangling interaction $S_y^2$ is contained in ${\cal U}_0$ in Eq.~\eqref{eq:propagator-0}.
In addition to that, ${\cal U}_0$ also contains terms proportional to $a S_y$ and $a^\dagger S_y$ corresponding to the annihilation and creation of a phonon conditioned on the state of the qubits.
A perfectly coherent gate in lowest order in $\eta$ is thus realized if $\av{f_1}$ vanishes at the gate time $T$, and if $\Phi_0(T)$ coincides with the desired Rabi angle $\Phi_T$.
In order to realize a coherent gate to higher order in $\eta$, it is necessary to ensure that the additional factors ${\cal U}_{j}(T)$ in Eq.~\eqref{eq:product} reduce to the identity or contribute solely to the coherent interaction between the qubits.
For general driving patterns, however, the factors ${\cal U}_{j}$ contain processes of the form $a^{\dagger p}a^q S_y^r$ with scalar prefactors depending on the driving profiles $f_k$.
The requirement that each of these prefactors must vanish at $t=T$ defines a constraint that the driving profiles must satisfy.
Since the derivation of each constraint follows exactly the same pattern, we will sketch it here with the process $a^\dagger a S_y^2$ in $\mathcal{U}_2$ as an illustrative example.
The corresponding prefactor is $\eta^2\bigl(\frac12\av[\big]{f_2\av{f_2^*}- f_2^*\av{f_2}}-2i\Phi_0  \bigr)$.
The requirement that the process $a^\dagger a S_y^2$ must not contribute to the gate thus results in the constraint $\Im\av[\big]{f_2\av{f_2^*}}=2\Phi_0(T)$ at the gate time.
In a similar fashion, all processes contributing to the full propagator $\mathcal V_d(T)$ to any desired order $d$ can be taken into account~\footnote{See Supplemental Material at \ldots\ for full details of all conditions at third order.}.

These constraints are satisfied for a broad range of driving profiles that can be selected depending on experimental constraints, goals and capabilities.
In the following we will discuss two such profiles that solve the conditions to third and fourth orders, that is considering terms up to and including $\eta^3$ or $\eta^4$.
To third order, only two sidebands are necessary, each driven monochromatically close to resonance with $f_1(t) = \Omega\exp(2i\delta t)$ and $f_2(t) = \Omega\exp(i\delta t)$ for a gate time of $T=2\pi/\delta$, with $\Omega$ determined by the entangling condition $\mathcal V_d=\exp(i\Phi_TS_y^2)$.
For this specific driving profile this condition is $3\eta^2x^4 - (1 + \eta^2)x^2 + \Phi_T/\pi = 0$, with $x=\Omega/\delta$, and the following discussion is based on the smallest positive root of this equation to minimize power usage.

The extension to fourth order requires driving the third sideband, and the driving profiles
\begin{align}
f_1(t) &= \Omega\exp(5i\delta t)\,, \qquad f_3(t) = \sqrt{\frac{3}{5}}\ \Omega\exp(i\delta t)\,,\\
f_2(t) &= \frac{\Omega}{\sqrt{5}}\Bigl(2\exp(2i\delta t) + \frac{7}{5}\frac{\Omega}{\delta}\eta\exp(-7i\delta t)\Bigl)\,,
\end{align}
with bichromatic driving of the second sideband, are a valid solution to all constraints, with $\Omega$ determined by
\begin{equation}\label{eq:three-sideband-entangling}
    \frac{382}{1875}x^6 - \frac{56}{75}\biggl(2+\frac{1}{\eta^2}\biggl)x^4
    +\biggl(1 + \frac{2}{\eta^2} +\frac{2}{\eta^4}\biggr)x^2 =  \frac{5}{\pi\eta^4}\Phi_T\,.
\end{equation}

\begin{figure}
    \includegraphics{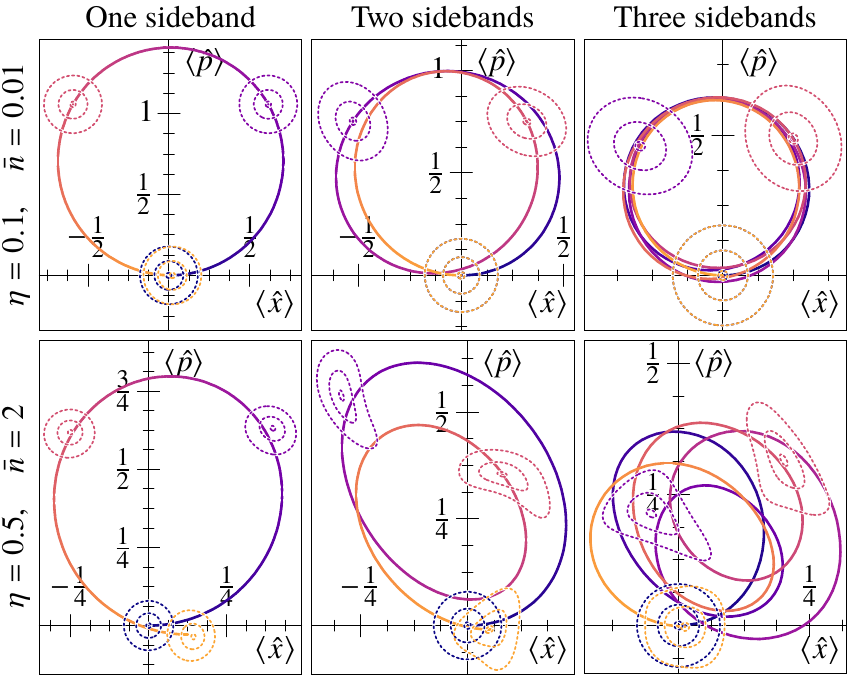}
    \caption{\label{fig:phasespace}%
Phase-space trajectories of gates using one, two and three sidebands (columns), both inside the Lamb--Dicke regime (top row) and far outside (bottom row).
Contours of the Wigner function of the motional state at times $0$, $T/3$, $2T/3$ and $T$ are indicated by the dashed lines, scaled down around their centroids to \SI{25}{\percent} (top) or \SI{5}{\percent} (bottom) for visibility.
Time through gate is represented by the color ranging from purple (dark) to orange (light).
A high-fidelity gate is realized only if the initial and final Wigner functions coincide.
Outside the Lamb--Dicke regime, increased contributions from the higher-order sidebands modify the phase-space trajectories and cause substantial deformation of the Wigner functions.
With conventional driving (one sideband) there is thus a substantial deviation between initial and final states.
With two sidebands, or even more with three, however, the phase-space trajectory closes rather accurately, and the distortion of the Wigner function vanishes to a good approximation at the end of the dynamics.}
\end{figure}

Figure~\ref{fig:phasespace} depicts the phase-space displacement of the motional mode during the gate operation for an initial state with the qubits in an eigenstate of $S_y$ and the motion in a thermal state, for conditions both inside (top) and outside (bottom) the Lamb--Dicke regime.
The left column illustrates the dynamics resulting from conventional driving of only first-order sidebands, while the central and right columns correspond to third- and fourth-order cases.
Outside the Lamb--Dicke regime, the phase-space trajectory (\textit{i.e.}\ the expectation of position and momentum) does not form a closed loop with the conventional driving scheme whereas the trajectories in the other two cases close near-perfectly as required for coherent quantum gates.

In addition to the phase-space trajectory, a scaled-down version of the Wigner function is depicted at thirds of the gate time with dashed contour lines.
With conventional driving (left) the contour shapes hardly change, but the present driving schemes result in strong deformations.
This is direct evidence of the second- and third-order sideband transitions that result in non-linear processes, and it is exactly those processes that manage to correct the phase-space trajectory.
Since the Wigner function and phase-space trajectory are not drawn to the same scale, the overlap between initial (orange, light) and final (purple, dark) Wigner functions is substantially larger than suggested by the figure.
While the figure therefore can not offer a quantitative estimate of this overlap, one can see that it becomes larger from left to right as expected, and the corresponding gate fidelities are \SI{71}{\percent}, \SI{89}{\percent} and \SI{97.3}{\percent}.

\begin{figure}
    \includegraphics{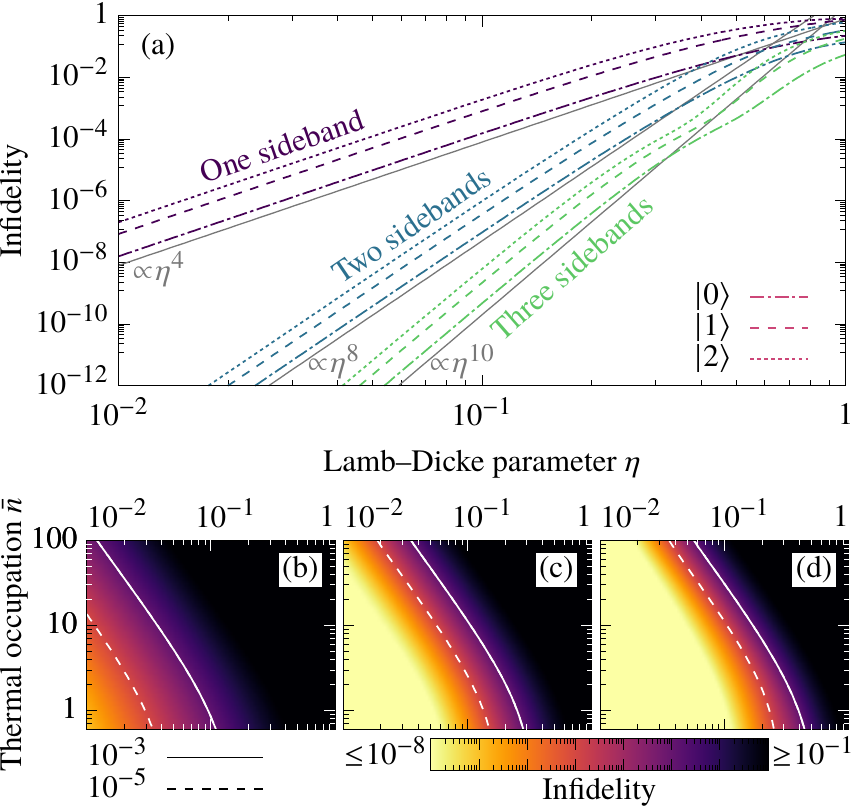}
    \caption{\label{fig:fidelity}%
(a) Gate infidelity as a function of Lamb--Dicke parameter $\eta$ for the schemes with one (purple, dark), two (blue, middle) and three (green, light) sidebands, starting from motional pure states $\ket0$ (dot-dashed), $\ket1$ (dashed) and $\ket2$ (dotted).
The present driving reduces the dependence of $\eta$ from $ O(\eta^4)$ for the standard gate to $\mathcal O(\eta^8)$ and $\mathcal O(\eta^{10})$ for the two- and three-sideband gates respectively.
Lines with the exact power laws are shown in solid gray for comparison.\\
(b)--(d) Heatmaps of the infidelity for a motional thermal state with varying mean occupation $\bar n$ for different values of $\eta$, using a scheme with (b) one, (c) two and (d) three sidebands.
Contours are plotted at infidelities of \num{e-3} (solid) and \num{e-5} (dashed).}
\end{figure}

The gate infidelity is strongly dependent on both the Lamb--Dicke parameter and the initial motional excitation, and is plotted as a function of the Lamb--Dicke parameter in Fig.~\ref{fig:fidelity}a for the three schemes under discussion, with different starting motional Fock states.
The solutions have the dependences ${\sim}\eta^4$, ${\sim}\eta^8$ and ${\sim}\eta^{10}$ that are consistent with the perturbative construction of the driving patterns.
In Figs.~\ref{fig:fidelity}b--d, this infidelity is plotted for each scheme respectively for the motion in an initial thermal state with varying mean occupation.
Extremely high fidelities can be reached for a broad range of Lamb--Dicke parameters in (c) and (d), whereas the conventional driving depicted in (b) requires a small value of the Lamb--Dicke parameter and a low motional temperature for a good gate fidelity.

Previous experimental work seeking fast gates has used a Lamb--Dicke parameter of approximately \num{0.1} with the average motional excitation cooled to $\bar n\lesssim0.05$~\cite{Schaefer18}.
At this level, outside-Lamb--Dicke effects lower-bound the single-sideband gate infidelity to \num{1.9e-4}, whereas the addition of a second and third sideband reduce the bound to \num{1.1e-7} and \num{7e-10} respectively.
The same fidelity as the base gate can be achieved by the second-order (third-order) scheme with a Lamb--Dicke parameter up to \num{0.27} (\num{0.43}) or a thermal state with average occupation $\bar n \le \num{6.6}$ (\num{21}).
That is, the present driving schemes would allow the realization of gates with state-of-the-art fidelities even without sideband cooling, and since all driving frequencies are spectrally close to the qubit transition frequency, both schemes can readily be realized with standard pulse-shaping equipment.

These two driving schemes illustrate high-fidelity gate operation with large Lamb--Dicke parameters and high temperatures of the motional mode, but the developed framework admits a broad spectrum of different solutions.
The sidebands may, for example, be driven polychromatically or with phase modulation to achieve robustness against heating and uncertainty in driving parameters~\cite{1367-2630-18-12-123007,PhysRevLett.121.180501,Shapira19,Milne2020}.
The techniques developed here thus should not be seen as an alternative to existing control techniques, but rather as a seamless addition to the existing ion-trapping toolbox.
While the general framework is exemplified here for the M\o lmer--S\o rensen scheme, it also readily applies to any other quantum gate that is based on a linearized interaction.
It is thus by no means limited to trapped ions, but can find application in any system for which fundamentally non-linear interactions are approximated to their leading order.
Prominent examples of such include superconducting qubits~\cite{Wendin17}, and optomechanical systems~\cite{RevModPhys.86.1391}.
As such, the present framework is not limited to the realization of quantum gates, but can also be applied for the design of high-precision sensors or other quantum-technological devices.

\begin{acknowledgments}
We are grateful for stimulating discussions with Joe Goodwin, Oliver Corfield, Richard Thompson and Simon Webster.
Financial support by EPSRC through the grant \textit{Optimal control for robust ion trap quantum logic} EP/P024890/1, and the \textit{Training and Skills Hub in Quantum Systems Engineering} EP/P510257/1 is gratefully acknowledged.
\end{acknowledgments}

\bibliography{generalized_gates_bib}

\begin{thebibliography}{25}%
\makeatletter
\providecommand \@ifxundefined [1]{%
 \@ifx{#1\undefined}
}%
\providecommand \@ifnum [1]{%
 \ifnum #1\expandafter \@firstoftwo
 \else \expandafter \@secondoftwo
 \fi
}%
\providecommand \@ifx [1]{%
 \ifx #1\expandafter \@firstoftwo
 \else \expandafter \@secondoftwo
 \fi
}%
\providecommand \natexlab [1]{#1}%
\providecommand \enquote  [1]{``#1''}%
\providecommand \bibnamefont  [1]{#1}%
\providecommand \bibfnamefont [1]{#1}%
\providecommand \citenamefont [1]{#1}%
\providecommand \href@noop [0]{\@secondoftwo}%
\providecommand \href [0]{\begingroup \@sanitize@url \@href}%
\providecommand \@href[1]{\@@startlink{#1}\@@href}%
\providecommand \@@href[1]{\endgroup#1\@@endlink}%
\providecommand \@sanitize@url [0]{\catcode `\\12\catcode `\$12\catcode
  `\&12\catcode `\#12\catcode `\^12\catcode `\_12\catcode `\%12\relax}%
\providecommand \@@startlink[1]{}%
\providecommand \@@endlink[0]{}%
\providecommand \url  [0]{\begingroup\@sanitize@url \@url }%
\providecommand \@url [1]{\endgroup\@href {#1}{\urlprefix }}%
\providecommand \urlprefix  [0]{URL }%
\providecommand \Eprint [0]{\href }%
\providecommand \doibase [0]{http://dx.doi.org/}%
\providecommand \selectlanguage [0]{\@gobble}%
\providecommand \bibinfo  [0]{\@secondoftwo}%
\providecommand \bibfield  [0]{\@secondoftwo}%
\providecommand \translation [1]{[#1]}%
\providecommand \BibitemOpen [0]{}%
\providecommand \bibitemStop [0]{}%
\providecommand \bibitemNoStop [0]{.\EOS\space}%
\providecommand \EOS [0]{\spacefactor3000\relax}%
\providecommand \BibitemShut  [1]{\csname bibitem#1\endcsname}%
\let\auto@bib@innerbib\@empty
\bibitem [{\citenamefont {Sch{\"a}fer}\ \emph {et~al.}(2018)\citenamefont
  {Sch{\"a}fer}, \citenamefont {Ballance}, \citenamefont {Thirumalai},
  \citenamefont {Stephenson}, \citenamefont {Ballance}, \citenamefont
  {Steane},\ and\ \citenamefont {Lucas}}]{Schaefer18}%
  \BibitemOpen
  \bibfield  {author} {\bibinfo {author} {\bibfnamefont {V.~M.}\ \bibnamefont
  {Sch{\"a}fer}}, \bibinfo {author} {\bibfnamefont {C.~J.}\ \bibnamefont
  {Ballance}}, \bibinfo {author} {\bibfnamefont {K.}~\bibnamefont
  {Thirumalai}}, \bibinfo {author} {\bibfnamefont {L.~J.}\ \bibnamefont
  {Stephenson}}, \bibinfo {author} {\bibfnamefont {T.~G.}\ \bibnamefont
  {Ballance}}, \bibinfo {author} {\bibfnamefont {A.~M.}\ \bibnamefont
  {Steane}}, \ and\ \bibinfo {author} {\bibfnamefont {D.~M.}\ \bibnamefont
  {Lucas}},\ }\href {https://doi.org/10.1038/nature25737} {\bibfield  {journal}
  {\bibinfo  {journal} {Nature}\ }\textbf {\bibinfo {volume} {555}},\ \bibinfo
  {pages} {75} (\bibinfo {year} {2018})}\BibitemShut {NoStop}%
\bibitem [{\citenamefont {Harty}\ \emph {et~al.}(2016)\citenamefont {Harty},
  \citenamefont {Sepiol}, \citenamefont {Allcock}, \citenamefont {Ballance},
  \citenamefont {Tarlton},\ and\ \citenamefont {Lucas}}]{Harty16}%
  \BibitemOpen
  \bibfield  {author} {\bibinfo {author} {\bibfnamefont {T.~P.}\ \bibnamefont
  {Harty}}, \bibinfo {author} {\bibfnamefont {M.~A.}\ \bibnamefont {Sepiol}},
  \bibinfo {author} {\bibfnamefont {D.~T.~C.}\ \bibnamefont {Allcock}},
  \bibinfo {author} {\bibfnamefont {C.~J.}\ \bibnamefont {Ballance}}, \bibinfo
  {author} {\bibfnamefont {J.~E.}\ \bibnamefont {Tarlton}}, \ and\ \bibinfo
  {author} {\bibfnamefont {D.~M.}\ \bibnamefont {Lucas}},\ }\href {\doibase
  10.1103/PhysRevLett.117.140501} {\bibfield  {journal} {\bibinfo  {journal}
  {Phys. Rev. Lett.}\ }\textbf {\bibinfo {volume} {117}},\ \bibinfo {pages}
  {140501} (\bibinfo {year} {2016})}\BibitemShut {NoStop}%
\bibitem [{\citenamefont {Gaebler}\ \emph {et~al.}(2016)\citenamefont
  {Gaebler}, \citenamefont {Tan}, \citenamefont {Lin}, \citenamefont {Wan},
  \citenamefont {Bowler}, \citenamefont {Keith}, \citenamefont {Glancy},
  \citenamefont {Coakley}, \citenamefont {Knill}, \citenamefont {Leibfried},\
  and\ \citenamefont {Wineland}}]{Gaebler16}%
  \BibitemOpen
  \bibfield  {author} {\bibinfo {author} {\bibfnamefont {J.~P.}\ \bibnamefont
  {Gaebler}}, \bibinfo {author} {\bibfnamefont {T.~R.}\ \bibnamefont {Tan}},
  \bibinfo {author} {\bibfnamefont {Y.}~\bibnamefont {Lin}}, \bibinfo {author}
  {\bibfnamefont {Y.}~\bibnamefont {Wan}}, \bibinfo {author} {\bibfnamefont
  {R.}~\bibnamefont {Bowler}}, \bibinfo {author} {\bibfnamefont {A.~C.}\
  \bibnamefont {Keith}}, \bibinfo {author} {\bibfnamefont {S.}~\bibnamefont
  {Glancy}}, \bibinfo {author} {\bibfnamefont {K.}~\bibnamefont {Coakley}},
  \bibinfo {author} {\bibfnamefont {E.}~\bibnamefont {Knill}}, \bibinfo
  {author} {\bibfnamefont {D.}~\bibnamefont {Leibfried}}, \ and\ \bibinfo
  {author} {\bibfnamefont {D.~J.}\ \bibnamefont {Wineland}},\ }\href {\doibase
  10.1103/PhysRevLett.117.060505} {\bibfield  {journal} {\bibinfo  {journal}
  {Phys. Rev. Lett.}\ }\textbf {\bibinfo {volume} {117}},\ \bibinfo {pages}
  {060505} (\bibinfo {year} {2016})}\BibitemShut {NoStop}%
\bibitem [{\citenamefont {Reagor~{\it et al}}(2018)}]{Reagor18}%
  \BibitemOpen
  \bibfield  {author} {\bibinfo {author} {\bibfnamefont {M.}~\bibnamefont
  {Reagor~{\it et al}}},\ }\href {\doibase 10.1126/sciadv.aao3603} {\bibfield
  {journal} {\bibinfo  {journal} {Science Advances}\ }\textbf {\bibinfo
  {volume} {4}} (\bibinfo {year} {2018}),\ 10.1126/sciadv.aao3603}\BibitemShut
  {NoStop}%
\bibitem [{\citenamefont {Veldhorst~{\it et al.}}(2015)}]{Veldhorst15}%
  \BibitemOpen
  \bibfield  {author} {\bibinfo {author} {\bibfnamefont {M.}~\bibnamefont
  {Veldhorst~{\it et al.}}},\ }\href {https://doi.org/10.1038/nature15263}
  {\bibfield  {journal} {\bibinfo  {journal} {Nature}\ }\textbf {\bibinfo
  {volume} {526}},\ \bibinfo {pages} {410} (\bibinfo {year}
  {2015})}\BibitemShut {NoStop}%
\bibitem [{\citenamefont {Rong}\ \emph {et~al.}(2015)\citenamefont {Rong},
  \citenamefont {Geng}, \citenamefont {Shi}, \citenamefont {Liu}, \citenamefont
  {Xu}, \citenamefont {Ma}, \citenamefont {Kong}, \citenamefont {Jiang},
  \citenamefont {Wu},\ and\ \citenamefont {Du}}]{Rong2015}%
  \BibitemOpen
  \bibfield  {author} {\bibinfo {author} {\bibfnamefont {X.}~\bibnamefont
  {Rong}}, \bibinfo {author} {\bibfnamefont {J.}~\bibnamefont {Geng}}, \bibinfo
  {author} {\bibfnamefont {F.}~\bibnamefont {Shi}}, \bibinfo {author}
  {\bibfnamefont {Y.}~\bibnamefont {Liu}}, \bibinfo {author} {\bibfnamefont
  {K.}~\bibnamefont {Xu}}, \bibinfo {author} {\bibfnamefont {W.}~\bibnamefont
  {Ma}}, \bibinfo {author} {\bibfnamefont {F.}~\bibnamefont {Kong}}, \bibinfo
  {author} {\bibfnamefont {Z.}~\bibnamefont {Jiang}}, \bibinfo {author}
  {\bibfnamefont {Y.}~\bibnamefont {Wu}}, \ and\ \bibinfo {author}
  {\bibfnamefont {J.}~\bibnamefont {Du}},\ }\href {\doibase 10.1038/ncomms9748}
  {\bibfield  {journal} {\bibinfo  {journal} {Nature Communications}\ }\textbf
  {\bibinfo {volume} {6}},\ \bibinfo {pages} {8748} (\bibinfo {year}
  {2015})}\BibitemShut {NoStop}%
\bibitem [{\citenamefont {Bermudez~{\it et al.}}(2017)}]{Bermudez17}%
  \BibitemOpen
  \bibfield  {author} {\bibinfo {author} {\bibfnamefont {A.}~\bibnamefont
  {Bermudez~{\it et al.}}},\ }\href {\doibase 10.1103/PhysRevX.7.041061}
  {\bibfield  {journal} {\bibinfo  {journal} {Phys. Rev. X}\ }\textbf {\bibinfo
  {volume} {7}},\ \bibinfo {pages} {041061} (\bibinfo {year}
  {2017})}\BibitemShut {NoStop}%
\bibitem [{\citenamefont {Bruzewicz}\ \emph {et~al.}(2019)\citenamefont
  {Bruzewicz}, \citenamefont {Chiaverini}, \citenamefont {McConnell},\ and\
  \citenamefont {Sage}}]{Bruzewicz19}%
  \BibitemOpen
  \bibfield  {author} {\bibinfo {author} {\bibfnamefont {C.~D.}\ \bibnamefont
  {Bruzewicz}}, \bibinfo {author} {\bibfnamefont {J.}~\bibnamefont
  {Chiaverini}}, \bibinfo {author} {\bibfnamefont {R.}~\bibnamefont
  {McConnell}}, \ and\ \bibinfo {author} {\bibfnamefont {J.~M.}\ \bibnamefont
  {Sage}},\ }\href {\doibase 10.1063/1.5088164} {\bibfield  {journal} {\bibinfo
   {journal} {Applied Physics Reviews}\ }\textbf {\bibinfo {volume} {6}},\
  \bibinfo {pages} {021314} (\bibinfo {year} {2019})}\BibitemShut {NoStop}%
\bibitem [{\citenamefont {Leibfried}\ \emph {et~al.}(2003)\citenamefont
  {Leibfried}, \citenamefont {Blatt}, \citenamefont {Monroe},\ and\
  \citenamefont {Wineland}}]{Leibfried03}%
  \BibitemOpen
  \bibfield  {author} {\bibinfo {author} {\bibfnamefont {D.}~\bibnamefont
  {Leibfried}}, \bibinfo {author} {\bibfnamefont {R.}~\bibnamefont {Blatt}},
  \bibinfo {author} {\bibfnamefont {C.}~\bibnamefont {Monroe}}, \ and\ \bibinfo
  {author} {\bibfnamefont {D.}~\bibnamefont {Wineland}},\ }\href {\doibase
  10.1103/RevModPhys.75.281} {\bibfield  {journal} {\bibinfo  {journal} {Rev.
  Mod. Phys.}\ }\textbf {\bibinfo {volume} {75}},\ \bibinfo {pages} {281}
  (\bibinfo {year} {2003})}\BibitemShut {NoStop}%
\bibitem [{\citenamefont {S\o{}rensen}\ and\ \citenamefont
  {M\o{}lmer}(2000)}]{Sorensen00}%
  \BibitemOpen
  \bibfield  {author} {\bibinfo {author} {\bibfnamefont {A.}~\bibnamefont
  {S\o{}rensen}}\ and\ \bibinfo {author} {\bibfnamefont {K.}~\bibnamefont
  {M\o{}lmer}},\ }\href {\doibase 10.1103/PhysRevA.62.022311} {\bibfield
  {journal} {\bibinfo  {journal} {Phys. Rev. A}\ }\textbf {\bibinfo {volume}
  {62}},\ \bibinfo {pages} {022311} (\bibinfo {year} {2000})}\BibitemShut
  {NoStop}%
\bibitem [{\citenamefont {Garc\'{\i}a-Ripoll}\ \emph
  {et~al.}(2005)\citenamefont {Garc\'{\i}a-Ripoll}, \citenamefont {Zoller},\
  and\ \citenamefont {Cirac}}]{Garcia05}%
  \BibitemOpen
  \bibfield  {author} {\bibinfo {author} {\bibfnamefont {J.~J.}\ \bibnamefont
  {Garc\'{\i}a-Ripoll}}, \bibinfo {author} {\bibfnamefont {P.}~\bibnamefont
  {Zoller}}, \ and\ \bibinfo {author} {\bibfnamefont {J.~I.}\ \bibnamefont
  {Cirac}},\ }\href {\doibase 10.1103/PhysRevA.71.062309} {\bibfield  {journal}
  {\bibinfo  {journal} {Phys. Rev. A}\ }\textbf {\bibinfo {volume} {71}},\
  \bibinfo {pages} {062309} (\bibinfo {year} {2005})}\BibitemShut {NoStop}%
\bibitem [{\citenamefont {Ospelkaus}\ \emph {et~al.}(2008)\citenamefont
  {Ospelkaus}, \citenamefont {Langer}, \citenamefont {Amini}, \citenamefont
  {Brown}, \citenamefont {Leibfried},\ and\ \citenamefont
  {Wineland}}]{Ospelkaus08}%
  \BibitemOpen
  \bibfield  {author} {\bibinfo {author} {\bibfnamefont {C.}~\bibnamefont
  {Ospelkaus}}, \bibinfo {author} {\bibfnamefont {C.~E.}\ \bibnamefont
  {Langer}}, \bibinfo {author} {\bibfnamefont {J.~M.}\ \bibnamefont {Amini}},
  \bibinfo {author} {\bibfnamefont {K.~R.}\ \bibnamefont {Brown}}, \bibinfo
  {author} {\bibfnamefont {D.}~\bibnamefont {Leibfried}}, \ and\ \bibinfo
  {author} {\bibfnamefont {D.~J.}\ \bibnamefont {Wineland}},\ }\href {\doibase
  10.1103/PhysRevLett.101.090502} {\bibfield  {journal} {\bibinfo  {journal}
  {Phys. Rev. Lett.}\ }\textbf {\bibinfo {volume} {101}},\ \bibinfo {pages}
  {090502} (\bibinfo {year} {2008})}\BibitemShut {NoStop}%
\bibitem [{\citenamefont {Bermudez}\ \emph {et~al.}(2012)\citenamefont
  {Bermudez}, \citenamefont {Schmidt}, \citenamefont {Plenio},\ and\
  \citenamefont {Retzker}}]{Bermudez12}%
  \BibitemOpen
  \bibfield  {author} {\bibinfo {author} {\bibfnamefont {A.}~\bibnamefont
  {Bermudez}}, \bibinfo {author} {\bibfnamefont {P.~O.}\ \bibnamefont
  {Schmidt}}, \bibinfo {author} {\bibfnamefont {M.~B.}\ \bibnamefont {Plenio}},
  \ and\ \bibinfo {author} {\bibfnamefont {A.}~\bibnamefont {Retzker}},\ }\href
  {\doibase 10.1103/PhysRevA.85.040302} {\bibfield  {journal} {\bibinfo
  {journal} {Phys. Rev. A}\ }\textbf {\bibinfo {volume} {85}},\ \bibinfo
  {pages} {040302} (\bibinfo {year} {2012})}\BibitemShut {NoStop}%
\bibitem [{\citenamefont {Steane}\ \emph {et~al.}(2000)\citenamefont {Steane},
  \citenamefont {Roos}, \citenamefont {Stevens}, \citenamefont {Mundt},
  \citenamefont {Leibfried}, \citenamefont {Schmidt-Kaler},\ and\ \citenamefont
  {Blatt}}]{Steane00}%
  \BibitemOpen
  \bibfield  {author} {\bibinfo {author} {\bibfnamefont {A.}~\bibnamefont
  {Steane}}, \bibinfo {author} {\bibfnamefont {C.~F.}\ \bibnamefont {Roos}},
  \bibinfo {author} {\bibfnamefont {D.}~\bibnamefont {Stevens}}, \bibinfo
  {author} {\bibfnamefont {A.}~\bibnamefont {Mundt}}, \bibinfo {author}
  {\bibfnamefont {D.}~\bibnamefont {Leibfried}}, \bibinfo {author}
  {\bibfnamefont {F.}~\bibnamefont {Schmidt-Kaler}}, \ and\ \bibinfo {author}
  {\bibfnamefont {R.}~\bibnamefont {Blatt}},\ }\href {\doibase
  10.1103/PhysRevA.62.042305} {\bibfield  {journal} {\bibinfo  {journal} {Phys.
  Rev. A}\ }\textbf {\bibinfo {volume} {62}},\ \bibinfo {pages} {042305}
  (\bibinfo {year} {2000})}\BibitemShut {NoStop}%
\bibitem [{\citenamefont {S\o{}rensen}\ and\ \citenamefont
  {M\o{}lmer}(1999)}]{Sorensen99}%
  \BibitemOpen
  \bibfield  {author} {\bibinfo {author} {\bibfnamefont {A.}~\bibnamefont
  {S\o{}rensen}}\ and\ \bibinfo {author} {\bibfnamefont {K.}~\bibnamefont
  {M\o{}lmer}},\ }\href {\doibase 10.1103/PhysRevLett.82.1971} {\bibfield
  {journal} {\bibinfo  {journal} {Phys. Rev. Lett.}\ }\textbf {\bibinfo
  {volume} {82}},\ \bibinfo {pages} {1971} (\bibinfo {year}
  {1999})}\BibitemShut {NoStop}%
\bibitem [{\citenamefont {Roos}(2008)}]{Roos08}%
  \BibitemOpen
  \bibfield  {author} {\bibinfo {author} {\bibfnamefont {C.~F.}\ \bibnamefont
  {Roos}},\ }\href {\doibase 10.1088/1367-2630/10/1/013002} {\bibfield
  {journal} {\bibinfo  {journal} {New Journal of Physics}\ }\textbf {\bibinfo
  {volume} {10}},\ \bibinfo {pages} {013002} (\bibinfo {year}
  {2008})}\BibitemShut {NoStop}%
\bibitem [{\citenamefont {Cohen}\ \emph {et~al.}(2015)\citenamefont {Cohen},
  \citenamefont {Weidt}, \citenamefont {Hensinger},\ and\ \citenamefont
  {Retzker}}]{Cohen15}%
  \BibitemOpen
  \bibfield  {author} {\bibinfo {author} {\bibfnamefont {I.}~\bibnamefont
  {Cohen}}, \bibinfo {author} {\bibfnamefont {S.}~\bibnamefont {Weidt}},
  \bibinfo {author} {\bibfnamefont {W.~K.}\ \bibnamefont {Hensinger}}, \ and\
  \bibinfo {author} {\bibfnamefont {A.}~\bibnamefont {Retzker}},\ }\href
  {\doibase 10.1088/1367-2630/17/4/043008} {\bibfield  {journal} {\bibinfo
  {journal} {New Journal of Physics}\ }\textbf {\bibinfo {volume} {17}},\
  \bibinfo {pages} {043008} (\bibinfo {year} {2015})}\BibitemShut {NoStop}%
\bibitem [{\citenamefont {Haljan}\ \emph {et~al.}(2005)\citenamefont {Haljan},
  \citenamefont {Brickman}, \citenamefont {Deslauriers}, \citenamefont {Lee},\
  and\ \citenamefont {Monroe}}]{Haljan05}%
  \BibitemOpen
  \bibfield  {author} {\bibinfo {author} {\bibfnamefont {P.~C.}\ \bibnamefont
  {Haljan}}, \bibinfo {author} {\bibfnamefont {K.-A.}\ \bibnamefont
  {Brickman}}, \bibinfo {author} {\bibfnamefont {L.}~\bibnamefont
  {Deslauriers}}, \bibinfo {author} {\bibfnamefont {P.~J.}\ \bibnamefont
  {Lee}}, \ and\ \bibinfo {author} {\bibfnamefont {C.}~\bibnamefont {Monroe}},\
  }\href {\doibase 10.1103/PhysRevLett.94.153602} {\bibfield  {journal}
  {\bibinfo  {journal} {Phys. Rev. Lett.}\ }\textbf {\bibinfo {volume} {94}},\
  \bibinfo {pages} {153602} (\bibinfo {year} {2005})}\BibitemShut {NoStop}%
\bibitem [{Note1()}]{Note1}%
  \BibitemOpen
  \bibinfo {note} {See Supplemental Material at \protect \ldots \ for full
  details of all conditions at third order.}\BibitemShut {Stop}%
\bibitem [{\citenamefont {Haddadfarshi}\ and\ \citenamefont
  {Mintert}(2016)}]{1367-2630-18-12-123007}%
  \BibitemOpen
  \bibfield  {author} {\bibinfo {author} {\bibfnamefont {F.}~\bibnamefont
  {Haddadfarshi}}\ and\ \bibinfo {author} {\bibfnamefont {F.}~\bibnamefont
  {Mintert}},\ }\href {http://stacks.iop.org/1367-2630/18/i=12/a=123007}
  {\bibfield  {journal} {\bibinfo  {journal} {N. J. Phys.}\ }\textbf {\bibinfo
  {volume} {18}},\ \bibinfo {pages} {123007} (\bibinfo {year}
  {2016})}\BibitemShut {NoStop}%
\bibitem [{\citenamefont {Webb}\ \emph {et~al.}(2018)\citenamefont {Webb},
  \citenamefont {Webster}, \citenamefont {Collingbourne}, \citenamefont
  {Bretaud}, \citenamefont {Lawrence}, \citenamefont {Weidt}, \citenamefont
  {Mintert},\ and\ \citenamefont {Hensinger}}]{PhysRevLett.121.180501}%
  \BibitemOpen
  \bibfield  {author} {\bibinfo {author} {\bibfnamefont {A.~E.}\ \bibnamefont
  {Webb}}, \bibinfo {author} {\bibfnamefont {S.~C.}\ \bibnamefont {Webster}},
  \bibinfo {author} {\bibfnamefont {S.}~\bibnamefont {Collingbourne}}, \bibinfo
  {author} {\bibfnamefont {D.}~\bibnamefont {Bretaud}}, \bibinfo {author}
  {\bibfnamefont {A.~M.}\ \bibnamefont {Lawrence}}, \bibinfo {author}
  {\bibfnamefont {S.}~\bibnamefont {Weidt}}, \bibinfo {author} {\bibfnamefont
  {F.}~\bibnamefont {Mintert}}, \ and\ \bibinfo {author} {\bibfnamefont
  {W.~K.}\ \bibnamefont {Hensinger}},\ }\href {\doibase
  10.1103/PhysRevLett.121.180501} {\bibfield  {journal} {\bibinfo  {journal}
  {Phys. Rev. Lett.}\ }\textbf {\bibinfo {volume} {121}},\ \bibinfo {pages}
  {180501} (\bibinfo {year} {2018})}\BibitemShut {NoStop}%
\bibitem [{\citenamefont {Shapira}\ \emph {et~al.}(2018)\citenamefont
  {Shapira}, \citenamefont {Shaniv}, \citenamefont {Manovitz}, \citenamefont
  {Akerman},\ and\ \citenamefont {Ozeri}}]{Shapira19}%
  \BibitemOpen
  \bibfield  {author} {\bibinfo {author} {\bibfnamefont {Y.}~\bibnamefont
  {Shapira}}, \bibinfo {author} {\bibfnamefont {R.}~\bibnamefont {Shaniv}},
  \bibinfo {author} {\bibfnamefont {T.}~\bibnamefont {Manovitz}}, \bibinfo
  {author} {\bibfnamefont {N.}~\bibnamefont {Akerman}}, \ and\ \bibinfo
  {author} {\bibfnamefont {R.}~\bibnamefont {Ozeri}},\ }\href {\doibase
  10.1103/PhysRevLett.121.180502} {\bibfield  {journal} {\bibinfo  {journal}
  {Phys. Rev. Lett.}\ }\textbf {\bibinfo {volume} {121}},\ \bibinfo {pages}
  {180502} (\bibinfo {year} {2018})}\BibitemShut {NoStop}%
\bibitem [{\citenamefont {Milne}\ \emph {et~al.}(2020)\citenamefont {Milne},
  \citenamefont {Edmunds}, \citenamefont {Hempel}, \citenamefont {Roy},
  \citenamefont {Mavadia},\ and\ \citenamefont {Biercuk}}]{Milne2020}%
  \BibitemOpen
  \bibfield  {author} {\bibinfo {author} {\bibfnamefont {A.~R.}\ \bibnamefont
  {Milne}}, \bibinfo {author} {\bibfnamefont {C.~L.}\ \bibnamefont {Edmunds}},
  \bibinfo {author} {\bibfnamefont {C.}~\bibnamefont {Hempel}}, \bibinfo
  {author} {\bibfnamefont {F.}~\bibnamefont {Roy}}, \bibinfo {author}
  {\bibfnamefont {S.}~\bibnamefont {Mavadia}}, \ and\ \bibinfo {author}
  {\bibfnamefont {M.~J.}\ \bibnamefont {Biercuk}},\ }\href {\doibase
  10.1103/PhysRevApplied.13.024022} {\bibfield  {journal} {\bibinfo  {journal}
  {Physical Review Applied}\ }\textbf {\bibinfo {volume} {13}},\ \bibinfo
  {pages} {024022} (\bibinfo {year} {2020})},\ \Eprint
  {http://arxiv.org/abs/1808.10462} {arXiv:1808.10462} \BibitemShut {NoStop}%
\bibitem [{\citenamefont {Wendin}(2017)}]{Wendin17}%
  \BibitemOpen
  \bibfield  {author} {\bibinfo {author} {\bibfnamefont {G.}~\bibnamefont
  {Wendin}},\ }\href {\doibase 10.1088/1361-6633/aa7e1a} {\bibfield  {journal}
  {\bibinfo  {journal} {Reports on Progress in Physics}\ }\textbf {\bibinfo
  {volume} {80}},\ \bibinfo {pages} {106001} (\bibinfo {year}
  {2017})}\BibitemShut {NoStop}%
\bibitem [{\citenamefont {Aspelmeyer}\ \emph {et~al.}(2014)\citenamefont
  {Aspelmeyer}, \citenamefont {Kippenberg},\ and\ \citenamefont
  {Marquardt}}]{RevModPhys.86.1391}%
  \BibitemOpen
  \bibfield  {author} {\bibinfo {author} {\bibfnamefont {M.}~\bibnamefont
  {Aspelmeyer}}, \bibinfo {author} {\bibfnamefont {T.~J.}\ \bibnamefont
  {Kippenberg}}, \ and\ \bibinfo {author} {\bibfnamefont {F.}~\bibnamefont
  {Marquardt}},\ }\href {\doibase 10.1103/RevModPhys.86.1391} {\bibfield
  {journal} {\bibinfo  {journal} {Rev. Mod. Phys.}\ }\textbf {\bibinfo {volume}
  {86}},\ \bibinfo {pages} {1391} (\bibinfo {year} {2014})}\BibitemShut
  {NoStop}%
\end{thebibliography}%
\end{document}


\title{Supplementary material to Strong-coupling quantum logic of trapped ions}
\author{Mahdi Sameti}
\affiliation{Blackett Laboratory, Imperial College London, London SW7 2AZ, United Kingdom}
\author{Jake Lishman}
\affiliation{Blackett Laboratory, Imperial College London, London SW7 2AZ, United Kingdom}
\author{Florian Mintert}
\affiliation{Blackett Laboratory, Imperial College London, London SW7 2AZ, United Kingdom}

\date{\today}

\begin{abstract}
This supplementary material contains the explicit conditions to be satisfied by driving profiles.
\end{abstract}

\maketitle

The desired entangling interaction that is accurate up to and including terms of order $\eta^3$ can be realized by driving only two sidebands with amplitudes $f_1$ and $f_2$.
In order to ease notation, the conditions that need to be satisfied are expressed in terms of the shorthand notations
\begin{equation}
\av{f} = \int_0^t\mathrm{d}t_1\,f(t_1)\,,
\end{equation}
where nesting implies iteration of the integration, {\it i.e.}
\begin{equation}
\av[\big]{f_1\av{f_2}} = \int_0^t\mathrm{d}t_1\,f_1(t_1)\int_0^{t_1}\mathrm{d}t_2\,f_2(t_2)\,
\end{equation}
and so on, and
\begin{equation}
\lambda(x,y) = x\av{y}-y\av{x}\,.
\end{equation}
The entangling condition at the gate time $t=T$ is
\begin{equation}\label{eq:third-order-entangling-functional}
\Phi_T = \Im\av*{f_1 \av{f_1^*} + \frac12\eta^2f_2\av{f_2^*}
                    - 4\eta^2 \av{f_1}\bigl(f_2^* \av[\big]{\av{f_1^\ast} f_2} + f_1 \av{f_1^\ast}^{2}\bigr)}\,.
\end{equation}
In addition to Eq.~\eqref{eq:third-order-entangling-functional}, all the following quantities also need to vanish at $t=T$: 
{\addtolength\jot{2\jot}\allowdisplaybreaks\begin{align}
    &\av{f_1}; \qquad \label{eq:c1} 
     \av{f_2}; \qquad
     \av[\big]{\av{f_1} f_2^*}; \qquad
     \av*{f_1 \av{f_1}}; \qquad
     \av[\big]{f_1\av{f_2^\ast}} ;\qquad
     \!\Re\av[\big]{\av{f_1} \av{f_1} f_2^*}; \qquad
     \av*{3f_1\av{f_2} + \av{f_1}f_2}; \\
    &\av*{\av{f_1^\ast}\bigl(2f_1^*\av{f_1} + f_2\av{f_2^\ast} - f_1\av{f_1^\ast}\bigr) - f_2^*\av[\big]{\av{f_1^\ast}f_2}}; \qquad
     \av*{2\av{f_2}\bigl(3\lambda(f_1, f_1^*) - \lambda(f_2, f_2^*)\bigr)
          + 3f_1\av[\big]{\av{f_1^\ast}f_2} + 3\av{f_1} \av{f_1^\ast}f_2};\\
    &\!\begin{aligned}
        \av*{12\av{f_1^\ast}\Re\bigl[f_1\av[\big]{\av{f_1} \av{f_2^\ast}}\bigr] + 6f_1\av{f_1} \av{f_1^*}\av{f_2^*} -4f_2\av{f_2^\ast}\av[\big]{\av{f_1}f_2^*} -3\av{f_1}^2\bigl(\av{f_1^\ast}f_2^* + f_1^*\av{f_2^*}\bigr)}\hspace*{8.6em}&\\[-\jot]
        {}+\av[\Big]{2\av{f_1}f_2^*\av{f_2}\av{f_2^\ast} - \av{f_1^\ast}^{3}\av{f_2} - 2\av[\big]{\av{f_1} f_2^*}\bigl(3f_1^*\av{f_1} - f_2^*\av{f_2}\bigr)}&;
        \end{aligned}\\
    &\!\Re\av*{
        4\av[\big]{\av{f_1}f_2^*}\Bigl(\av{f_1}\bigl(6f_1\av{f_1^\ast} - 3f_1^*\av{f_1} + 2f_2^*\av{f_2}\bigr) - 2f_2\av[\big]{\av{f_1}f_2^*}\Bigr)
        -4\av{f_1}^3\av{f_1^\ast}f_2^* + 3f_1\av{f_1}\av{f_2^\ast}\label{eq:c2}
        };\\
    &\av*{2\lambda(f_1, f_1^*) - \lambda(f_2, f_2^*)}. \label{eq:resonant}
\end{align}}%
All quantities listed in \crefrange{eq:c1}{eq:c2} are satisfied by the driving fields specified in the main article, because they are monochromatic with frequencies given by $\delta$ and $2\delta$.
\Cref{eq:resonant} is satisfied because in addition to this, $f_1(t)$ and $f_2(t)$ have the same magnitude.

\bigskip

The conditions for an entangling interaction that is accurate up to and including terms of order $\eta^4$ requires driving of three sidebands.
These conditions are excessively long, but can be found electronically at \url{https://www.github.com/ImperialCQD/Strong-Coupling-Quantum-Logic-of-Trapped-Ions}, where the conditions in \crefrange{eq:third-order-entangling-functional}{eq:resonant} are also given.